\documentstyle[pre,aps,psfig,floats,twocolumn]{revtex}
\begin{document}
%{\noindent \it Submitted to \small ?}
\draft
\twocolumn[\hsize\textwidth\columnwidth\hsize\csname 
@twocolumnfalse\endcsname

\title{Democracy versus Dictatorship in Self-Organized Models of Financial Markets}
\author{R. d'Hulst and G.J. Rodgers}
\address{Rene.DHulst@brunel.ac.uk and G.J.Rodgers@brunel.ac.uk}
\address{Department of Mathematical Sciences, Brunel University}
\address{Uxbridge, Middlesex, UB8 3PH, UK}

\maketitle
\begin{abstract}
Models to mimic the transmission of information in financial markets
are introduced. As an attempt to generate the demand process, we 
distinguish between dictatorship associations, where groups of agents 
rely on one of them to make decision, and democratic associations, 
where each agent takes part in the group decision. In the dictatorship 
model, agents segregate into two distinct populations, while the 
democratic model is driven towards a critical state where groups of 
agents of all sizes exist. Hence, both models display a level of 
organization, but only the democratic model is self-organized. We show 
that the dictatorship model generates less volatile markets than the 
democratic model.
\end{abstract}

\pacs{PACS: 89.90.+n, 02.50.Le, 64.60.Cn, 87.10.+e}
\pacs{Keywords: herding, economy, market organization, decision process.}

]
\narrowtext

\section{Introduction}
\label{sec:introduction}

Understanding the microscopic dynamics of financial markets is a challenging problem that has recently attracted the attention of physicists \cite{bouchaud,mantegna,vandewalle,vandewalle2}. There have been a lot of studies of the time series of returns on various time scales \cite{mandelbrot,stanley}. For intraday variations, L\'evy stable distributions with parameter between 1.35 and 1.8 have been proposed for the distribution of the returns \cite{stanley}, while on longer time scales, the distribution seems to converge towards a normal distribution. Moreover, the large fluctuations on short time scales are observed to decrease according to a power law with an exponent approximately equal to 3 \cite{stanley}.

The short time scale distribution for the returns is in obvious contradiction of one the first models for financial markets, in which it is assumed that the stochastic process for the returns is an uncorrelated random walk \cite{bachelier}. In such a framework, the distribution for the returns should converge towards a Gaussian distribution, according to the central limit theorem. Several scenarios have been proposed to account for the deviations from Gaussian distributions, in terms of weakly interacting agents \cite{mandelbrot}, multi-agent markets \cite{bak}, through heteroskedasticity \cite{arch} or through herding \cite{cont}. Each of these explanations has its own motivations and associated drawbacks. These are discussed and documented in \cite{cont}. In this paper, we concentrate on the hypothesis of herding, whose main advantage is that it is so simple that some analytical calculations can be carried out.\\
\indent\indent Herding assumes that agents are not making decisions independently, but that each agent is a member of a group that makes a collective decision. The larger this group is, the bigger the impact it has on the market when the group trades together. In the herding assumption, these groups are responsible for the large fluctuations observed in the changes of the prices. The existence of herding is well documented \cite{scharfstein,trueman,grinblatt}, while its origin can be interpreted in a number of ways, such as by agents sharing the same information, using the same analytical tools or following the same rumours.

In this paper, we propose two extensions of previous models for herding \cite{cont,eguiluz}. In these early models, no attempt was made to consider the generation of the demand process. Consequently, the number of transactions had to be tuned to a low level to ensure the appearance of fat tails in the change of price distribution. In this paper, the process of demand is modelled with two types of association, that we choose to call dictatorship and democratic. Both models are explained in detail in the next section, while they are investigated numerically and analytically in the later sections. 

\section{The Models}
\label{sec:the models}

The proposed models are extensions of the model of Egu\'{\i}luz and Zimmermann \cite{eguiluz,dhulst}, which is itself a dynamical formulation of a model by Cont and Bouchaud \cite{cont}. Initially, every agent is given a random number $p$ between 0 and 1, chosen from a uniform distribution. A cluster of agents is a set of agents that share the same value of $p$. Agents belonging to a cluster make the same decisions between buying, selling or doing nothing. At the beginning of the simulation, all the agents are independent, with just one agent per cluster. At each time step, two agents $i$ and $j$, with associated numbers $p_i$ and $p_j$ respectively, are selected at random. With a probability $a_{ij} = |p_i - p_j|$, agent $i$ and all the agent belonging to his cluster are given a new random number chosen from the range $\lbrack p_i-R, p_i+R \rbrack$. We say that all the links between these agents are removed or that the cluster of agent $i$ is fragmented. With a probability $1 - a_{ij}$, all the agents belonging to the clusters of agents $i$ and $j$ are given the same number $p_{ij}$. In other words, the two clusters coagulate. For the dictatorship model, $p_{ij} = p_i$, while for the democratic model, $p_{ij} = (p_i + p_j ) /2$.

We assume that $p$ is a simple parameter used to code the character of an agent. According to our definition of $a_{ij}$, agents with similar characters are more inclined to associate, that is for $p_i\simeq p_j$, $a_{ij}$ is close to 0. Of course, this terminology should not be taken literally. Apart from the highly unrealistic idea of coding somebody's character in just one real number, the dynamics of the models imply that agents are changing their characters by making associations. Consequently, even if we call $p$ the character of an agent, it would be better to keep in mind that $p$ reflects the way an agent is perceived in the market, rather than the personality of the agent himself. In the dictatorship model, all the agents belonging to a cluster are sharing the character of one of them. In other words, the whole cluster makes decisions according to the character of one agent. On the contrary, in the democratic model, the character of a cluster of agents is a linear combination of the characters of all the agents belonging to the cluster. This means that the character of each agent belonging to a cluster has an influence on the decision process. The chosen average in the democratic model is not strictly speaking democratic, because not all the agents have the same weight on the character of the cluster. However, various weighted averages have been considered, using the size of the clusters coagulating for instance, but no qualitative difference in the results was observed. As a consequence, we will stick to the previously defined average for its analytical simplicity. 

As in the original paper by Egu\'{\i}luz and Zimmermann \cite{eguiluz}, we associate the fragmentation of a cluster with the decision from the group of agents that it is time to buy or to sell, both decisions being equally likely to occur. After a transaction is performed, the links between the agents are removed. This process tends to mimic the dynamics of the connections in financial markets, where associations are usually for a fixed time, or for a given purpose. When the deadline or the purpose is attained, agents are free to renew their association or to try new contacts. Alternatively, if the association of agents is interpreted as agents sharing the same information, this information becomes useless after a transaction, and the association no longer exists. In our model, we simply give new values for the character $p$ of each agent in a given range $R$ around the old value $p_{0}$. This choice allows us to introduce a kind of memory into the system. Agents that were already connected are more likely to connect again, because their $p$ values are closer than two agents chosen at random.

To compare the two types of coagulation process, a price is defined. The price is generated by the difference between the supply and demand. Every time that a transaction is decided, the difference between the number of buying and selling orders is modified. We define the return ${\cal R}$ at a given time $t$ as the number of agents buying at that time. If there is no transaction, ${\cal R} (t) =0$, a cluster of $s$ agents buying gives ${\cal R} (t) = s$ and for a cluster of $s$ agents selling, ${\cal R} (t) = -s$. The change in the logarithm of the price is assumed to be proportional to ${\cal R}$  \cite{cont}, that is

\begin{equation}
\ln P (t) - \ln P (t-1) = {1\over \lambda} {\cal R} (t).
\label{eq:definition of the price}
\end{equation}
The logarithm is introduced to ensure that the price is always a positive quantity. The proportionality coefficient $\lambda$ determines the sensitivity of the price to variations in the demand. The probability to have a return ${\cal R}$ of size $s$ is equal to the probability of having a cluster of $s$ agents making a transaction. If $n_s$ is defined as the number of clusters of size $s$ for a system with $N_0$ agents, the probability of having ${\cal R}=s$ for a transaction is equal to $s n_s /N_0$.

\section{Numerical results for the democratic model}
\label{sec:results for the democratic model}

If $n_s$ is the number of clusters of size $s$, $s n_s / N_0$ is the probability to select an agent belonging to a cluster of size $s$ when selecting an agent at random. The distribution $n_s /N_0$ is presented in Fig. 1 for markets with $N_0 = 10^5$ agents, when the range $R$ is fixed to 0.1. The cluster size distribution can be approximated by a power law, as
\begin{figure}
\centerline{\psfig{file=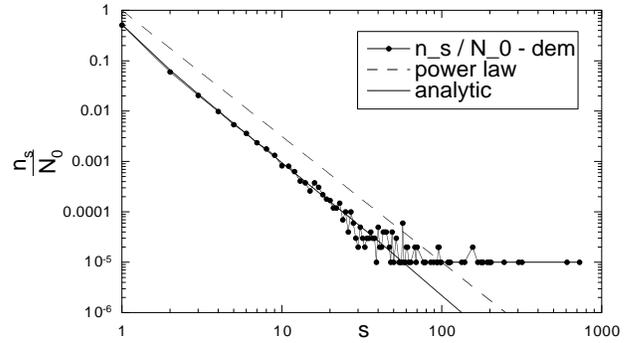,width=8.5cm}}
\caption{Cluster size distribution $n (s) / N_0$ for the democratic model for $N_0 = 10^5$ (continuous line with $\bullet$) agents after $t = 10^7$ time steps. The range is fixed to $R=0.1$. The dashed line is a guide to the eyes for a power law of exponent $-5/2$, while the continuous line is the analytical expression for $n_s / N_0$ for $\overline{a}= 0.1$.}
\label{fig:cluster size distribution-democratic}
\end{figure}

\begin{equation}
n_s \sim s^{- \alpha} f \left( {s\over s_c} \right)
\label{eq:scaling of ps distribution}
\end{equation}
where $s_c$ is the critical size where finite size effects become dominant. In the previous equation, $f (x) = 1$ for $x<1$ and is going to 0 otherwise. To obtain an estimate for $s_c$, we assume that $n_s / N_0 = (\alpha -2) s^{-\alpha}$. For a particular choice of value of $N_0$, the size effects should be very important for $s = s_c$ where $n_{s_c} \simeq 1$, giving $s_c \simeq (N_0 (\alpha -2))^{1/\alpha }$, in agreement with the numerical simulations. For the largest size investigated, that is, for $N_0 = 10^5$, we obtain $\alpha = 2.6\pm 0.1$. Our analytical results suggest that $\alpha = 5/2$ for $N_0\rightarrow \infty$, but with an exponential correction to the power law, as shown by the continuous line in Fig. 1. A dashed line corresponding to a power law of exponent $5/2$ is drawn for comparison. The range $R$ has no noticeable effect on $n_s$.  Numerically, we observe that the smaller size systems always reach a frozen state, with a unique cluster including all the agents. This is expected if the frequency of fragmentation is very low, as will be seen from the results for the distribution of the characters $p$ of the agents.

In Fig. 2, the distribution of the characters $p$ is shown for $N_0 = 10^5$ agents, $R=0.1$ and $R=0.01$, for a discretization of 50 points. $Q (p) dp$ is defined as the probability that an agent chosen at random has a character $p$ included in the range $\lbrack p, p + dp \rbrack$. As a first approximation, the distribution can be written
\begin{figure}
\centerline{\psfig{file=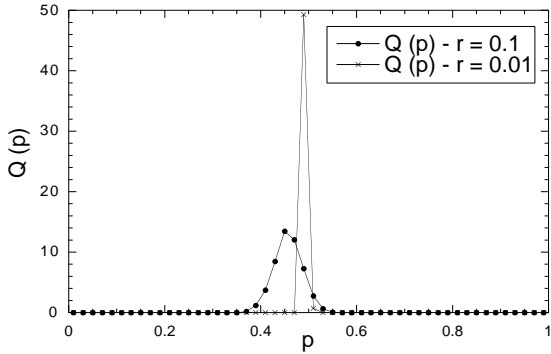,width=8.5cm}}
\caption{Probability distribution $Q (p)$ of the characters $p$ of the agents for the democratic model for $N_0 = 10^4$ agents after $t = 10^7$ time steps. The range is $R = 0.1$ ($\bullet$) and $R=0.01$ ($\times$).}
\label{fig:character distribution-democratic}
\end{figure}

\begin{equation}
Q (p) = \delta (p-p_0)
\label{eq:qp democratic}
\end{equation}
with the Dirac delta distribution $\delta (x)$. $p_0$ is a value of $p$ that is shared by most of the agents. As most of the agents have a value of $p$ close to $p_0$, the fragmentation rate is very low. The spreading around the value of $p_0$ is of the order of the range $R$, giving Eq. (\ref{eq:qp democratic}) as an exact solution in the limit $R\rightarrow 0$.

Starting from a uniform distribution, the stationary value for $p_0$ is around 0.5. However, the convergence to a unique value $p_0$ for most of the agents comes from the averaging process for the coagulation of clusters. As a result, $p_0$ is strongly dependent of our choice for the initial distribution. For instance, using the distribution

\begin{equation} 
S (p) = \left\{ 
\begin{array}{ll}
{1\over h}& \hbox{for }p<h \\
0& \hbox{otherwise}\\
\end{array}
\right.
\end{equation}
gives $p_0\simeq h/2$. The fragmentation process also generates a very slow drift in the value for $p_0$. Hence, the property of the democratic model is that most agents are associated with a value $p$ near a common value $p_0$, but the value of $p_0$ itself is meaningless. This property could have been anticipated since $a_{ij}$ only depends on the difference between $p_i$ and $p_j$, not on the absolute value of $p_i$ or $p_j$. 

We define an agent to be active at time $t$ if she is the one making her cluster buy or sell at time $t$. The distribution of the time between two successive activations of an agent has been investigated, but it was found to coincide with a random activation of one agent among $N_0$. Simulations of the model were also performed on one- and two-dimensional square lattices, with one agent corresponding to one site. Connections are restricted to first neighbouring sites. The distribution of the distance between two successive active agents was found to coincide with a random choice of an active site at each time step. We conclude that the model does not display any temporal or spatial correlations, apart from the clustering of agents.

\section{Numerical results for the dictatorship model}
\label{sec:numerical results for the dictatorship model}

The same numerical simulations have been carried out for the dictatorship model. Fig. 3 presents the size distribution $n_s / N_0$ for the dictatorship model for $N_0 = 10^5$ agents and a value $R = 0.1$ for the range. 
\begin{figure}
\centerline{\psfig{file=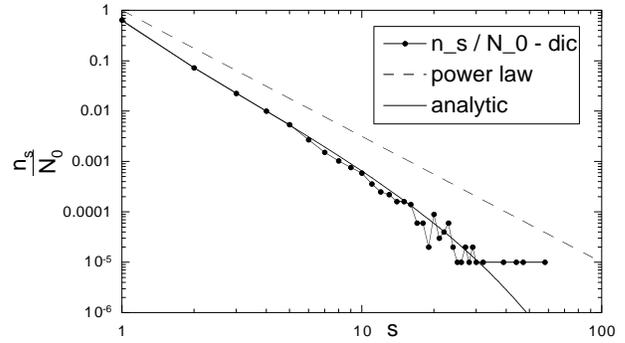,width=8.5cm}}
\caption{Cluster size distribution $n (s) / N_0$ for the dictatorship model for $N_0 = 10^5$ agents (continuous line with $\bullet$) after $t = 10^7$ time steps. The range is fixed to $R=0.1$. The dashed lines is a guide to the eyes for a power law of exponent $-5/2$, while the continuous line represents Eq. (\ref{eq:distribution of the clusters}) for $\overline{a} = 0.4$.}
\label{fig:cluster size distribution-dictatorship}
\end{figure}
The dictatorship model roughly displays a power law dependence for the cluster size distribution, with finite size effects becoming important for $s_c \simeq (N_0 (\alpha -2))^{1/\alpha }$, as given in the previous section. $s_c$ is slightly overestimated compared with numerical simulations. The exponent $\alpha$, as defined in Eq. (\ref{eq:scaling of ps distribution}), is estimated to be $\alpha = 3.0\pm 0.2$ for $N_0 = 10^5$ agents. A value of $\alpha = 5/2$ is expected from our analytical results, but with an exponential correction as represented by the continuous line in Fig. 3. The dashed line shows a power law of exponent 5/2 for comparison. $R$ does not have any noticeable effect on $n_s /N_0$.

The distribution of the characters of the agents, $Q (p)$, is very different in this model compared to the previous one. As can be seen in Fig. 4 obtained for $50$ points with the choices $N_0 = 10^4$ agents, $R = 0.1$ and $R = 0.01$, most agents are associated with values of $p$ close to 0 or 1. As a result, the total population of agents can be decomposed into two subpopulations, one associated with $p\simeq 0$ and one associated with $p\simeq 1$. Taking smaller values for $R$ enforces the separation in two populations, suggesting that 
\begin{figure}
\centerline{\psfig{file=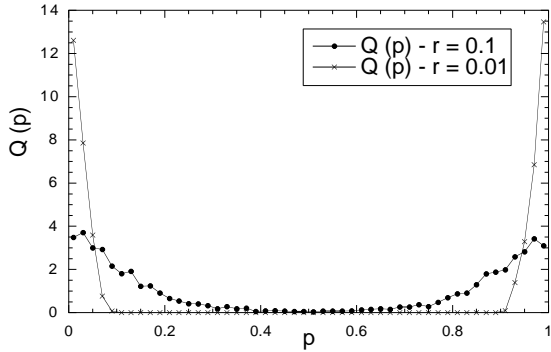,width=8.5cm}}
\caption{Probability distribution $Q (p)$ of the characters $p$ of the agents for the dictatorship model for $N_0 = 10^5$ agents after $t=10^7$ time steps. The range is $R = 0.1$ ($\bullet$) and $R=0.01$ ($\times$).}
\label{fig:character distribution-dictatorship}
\end{figure}

\begin{equation}
Q (p) = {\delta(p) + \delta (1-p)\over 2}
\label{eq:qp dictatorship} 
\end{equation}
when $R\rightarrow 0$. Agents belonging to the same population have a high probability to be connected when they interact. Conversely, agents belonging to different populations would prefer to make a transaction rather than communicate with each other. This gives an average fragmentation rate $\overline{a}$ close to 0.5, because the two populations are of similar sizes. As this binary distribution is not put in by hand, one is tempted to say that the model is self-organized, meaning that a complex pattern has appeared as a result of the dynamics of the model. However, as will be shown later, the model is not self-organized in the usual sense of a system being driven into a critical state. 

The distribution of the time between two successive activations of an agent has also been investigated, giving a random choice of one agent among $N_0$ at each time step. On-lattice simulations of the model show that the distribution of the distance between two successive active agents coincides with a random choice of an active site at each time step. The dictatorship model, like the democratic model, does not display any temporal or spatial correlations, apart from the clustering of agents.

\section{Analytical Results}
\label{sec:analytical results}

Both models exhibit stationary cluster size distributions that are similar to those obtained in the original models \cite{cont,eguiluz,dhulst}. In this section, we assume that the cluster size distribution $n_s /N_0$ and the character distribution $Q (p)$ are independent. The average fragmentation rate of a cluster of character $p$ is defined by

\begin{equation}
a (p) = \int_0^1 dp_1 Q (p_1) | p_1 - p|.
\label{eq:a as a function of p}
\end{equation}
We assume that $n_s$ is a function of the average fragmentation rate $\overline{a}$ only, defined by

\begin{equation}
\overline{a} = \int_0^1 dp Q (p) a (p).
\end{equation}
We have checked that the stationary cluster size distributions corresponding to different range of values for $p$ are equivalent to one another. The number $n_s$ of clusters of size $s>1$ evolves like

\begin{eqnarray}\nonumber 
{\partial n_s\over \partial t} &=& -\overline{a} s n_s + {(1-\overline{a})\over N_0}
\sum_{r=1}^{s-1} r n_r (s-r) n_{s-r} \\
&& - {2(1-\overline{a}) s n_s\over N_0} \sum_{r=1}^{\infty} r n_r
\label{eq:evolution of ns}
\end{eqnarray}
where $N_0$ is the total number of agents. Note that one time step in this
formulation corresponds to one attempted update per agent in the numerical
simulation. The first term on the right-hand side of Eq.~(\ref{eq:evolution of ns}) describes the fragmentation of a cluster of size $s$ in the case of a transaction involving $s$ agents. The second term describes the creation of a new cluster of size $s$ by coagulation of two clusters of size $r$ and $s-r$, with $r<s$. The last term on the right-hand side describes the disappearance of a cluster of size $s$ by coagulation with another cluster. The number of clusters of size 1, $n_1$ obeys

\begin{equation}
{\partial n_1\over \partial t} = \overline{a} \sum_{r=2}^{\infty} r^2 n_r - {2(1-\overline{a})
n_1\over N_0} \sum_{r=1}^{\infty} r n_r.
\label{eq:evolution of n1}
\end{equation}
The first term on the right-hand side describes the appearance of $s$
clusters of size 1 by fragmentation of a cluster of size $s$ while the second
term describes the disappearance of clusters of size 1 by coagulation with
another cluster. As shown in \cite{dhulst}, the cluster size distribution is given by 

\begin{equation}
n_s \sim N_0 \left({4 (1-\overline{a})\over (2-\overline{a})^2}\right)^s s^{-5/2}.
\label{eq:distribution of the clusters}
\end{equation}
Hence, the size distribution of the clusters follows a power law corresponding to $\alpha = 5/2$, with an exponential cut-off, as in the original models \cite{cont,dhulst}. For $\overline{a}= 0$, the exponential cut-off vanishes.

For the democratic model, $\overline{a}$ is close to 0 in the stationary state, all the agents being represented by a value of $p$ close to a common value $p_0$. We present Eq. (\ref{eq:distribution of the clusters}) for $\overline{a}=0.1$ in Fig. 1. This situation is reminiscent of intraday transactions on financial markets, where a value of $\overline{a}\rightarrow 0$ is expected. On the contrary, for the dictatorship model, $\overline{a}$ is close to 0.5 in the stationary state, because the agents segregate into two populations of comparable sizes. The probability of choosing two agents belonging to the same population is almost equal to the probability of choosing one agent in each population. For the former, a coagulation process will almost certainly take place, while for the latter a fragmentation process is very likely to happen. Hence, coagulation and fragmentation are equally likely to occur. Our analytical result for the cluster size distribution is in reasonable agreement with the results from the numerical simulations, as can be appreciated in Fig. 1 and 3 for the democratic and the dictatorship models, respectively. The numerical estimates of $\alpha$ for both models are larger than the analytical ones, which is to be expected as the exponential correction term is a decreasing function of $s$ for $\overline{a}\in ( 0,1)$. Also, because $R \not = 0$, $\overline{a}$ is lower than 1/2 in the dictatorship model. In Fig. 3, we present Eq. (\ref{eq:distribution of the clusters}) for $\overline{a}=0.4$.

Note that the two values of $\overline{a}$ obtained, $\overline{a}=0$ and $\overline{a}=0.5$, are the minimum and the maximum values of $\overline{a}$ that can be achieved in our models. For a character $p$ equal to a real number between 0 and 1, $\overline{a}$ can not be greater than 0.5. Relaxing this constraint should allow us to explore a richer set of models, while our two variations seems to be representative of this class of models. In fact, several variations for the coagulation process have been investigated and all of them qualitatively reproduced the results of either the democratic or the dictatorship model.   

We assumed in Sec. \ref{sec:the models} that the variation of the price is a function of the variation of the balance between the number of buyers and sellers through Eq. (\ref{eq:definition of the price}). For any transaction, the probability to have a return ${\cal R}$ of size $s$ is $s n_s/N_0$. As the fragmentation rate is lower for the democratic model compared to the dictatorship model, bigger clusters are more likely to form, giving bigger changes in the price. As a result, prices in the democratic model have a higher volatility.   

Translating our models into SOC language \cite{bak2}, a transaction can be associated with an avalanche. As all the agents taking part in the same transaction share the same character $p$, they are obviously correlated with each other, showing that all the elements taking part in an avalanche are correlated. As we showed numerically, no other spatial or temporal correlations were detected in either model. To have a system in a critical state, the average size of a transaction should diverge. As shown in \cite{dhulst}, the average size of a transaction is given by $1/\overline{a}$. Hence, the democratic model is self-organized but the dictatorship model is not. The two processes characteristic of a self-organized system are present, accumulation in the form of a coagulation process and relaxation by fragmentation. For $\overline{a}\rightarrow 0$, there is separation between the time scales of these two processes, the coagulation process appearing typically at each time step of the simulation, while the fragmentation process appears on average every $1/\overline{a}$ time steps. This separation of time scales is not present for any other values of $\overline{a}$, except for $\overline{a}\rightarrow 1$, where no correlation between the agents can be built up because the fragmentation rate is too high.

For the dictatorship model, the time master equation for $Q (p)$ is

\begin{eqnarray}
{\partial Q (p)\over \partial t} &=& Q (p) \int_0^1 dp_1 Q (p_1) (1 - |p-p_1|) \times\\
\label{eq:equilibrium qp dictatorship}
\nonumber
&\times& \left( {1\over a (p_1)} - {1\over a (p)} \right) + {R^2\over 8} {\partial^2 Q (p)\over \partial p^2}.
\end{eqnarray}
One time step corresponds to one attempted update per agent. The first term consists of two contributions. The coagulation of a cluster in $p$ and a cluster in $p_1$ to create a cluster in $p$ and the coagulation of a cluster in $p$ and a cluster in $p_1$ to create a cluster in $p_1$. Note that the average size of a cluster of character $p$ is equal to $1/ a (p)$ \cite{dhulst}. The last term corresponds to the diffusion of agents by cluster fragmentation.

As can be seen from numerical simulations, the segregation in two populations originates from the coagulation process and is enhanced when $R\rightarrow 0$. Hence, we neglect the diffusive term proportional to $R^2$ as a first approximation, giving the stationary state as

\begin{equation}
Q (p) \int_0^1 dp_1 Q (p_1) (1 - |p-p_1|)\left( {1\over a (p_1)} - {1\over a (p)} \right) = 0.
\end{equation}
A possible solution to this equation is $a (p) = \overline{a}$, independent of $p$. Taking the derivative of Eq. (\ref{eq:a as a function of p}) with respect to $p$, you have

\begin{equation}
- \int_p^1 dp_1 Q (p_1) + \int_0^p dp_1 Q (p_1) = 0  
\end{equation}
The previous relation has $Q (p) = (\delta (p) + \delta (1 - p) ) /2$ as solution. Eq. (\ref{eq:qp dictatorship}) is the stationary distribution for $R\rightarrow 0$.   

For the democratic model, the time master equation for the distribution $Q (p)$ is

\begin{eqnarray}
\nonumber
{\partial Q (p) \over \partial t} &=& - {Q (p)\over a (p)} \int_0^1 dp_1 Q (p_1) (1-|p-p_1|) + {R^2\over 8a} {\partial^2 Q (p)\over \partial p^2}\\
&+& \int_0^1 \int_0^1 dp_1 dp_2 Q (p_1) Q (p_2) {a (p_1) + a (p_2)\over a (p_1)a (p_2)} \times\\
\nonumber
&& \qquad\times (1 - |p_1 - p_2|) \delta \left({p_1+p_2-2p\over 2}\right).
\end{eqnarray}
The first term on the right-hand side describes the average decrease of $Q (p)$ when clusters in $p$ are coagulating with clusters of character $p_1$. The last term is the inverse process, when $Q (p)$ increases by coagulation of two clusters. The second term corresponds to the diffusion of agents due to the fragmentation of clusters. 

As shown by the numerical results, the diffusive process when a cluster is fragmented accounts for a very small drift of the previously introduced value of $p_0$. But it is not sufficient to make this value independent of initial conditions. Hence, we neglect the term proportional to $R^2$ as a first approximation. The stationary state is obtained for

\begin{eqnarray}
\nonumber &&
\int_0^1 dp_1 Q (p_1) (2 Q (2p-p_1) (1 - 2|p-p_1|)\\
&& - Q (p) (1 - |p-p_1|)) = 0,
\end{eqnarray}
This equation has $Q (p) = \delta (p-p_0)$ as solution, where $p_0$ is free to take any value. The final value for $p_0$ is a function of the initial choice for $Q (p)$ and because of the particular choice of averaging process, we have

\begin{equation}
p_0 \simeq \int_0^1 p\ Q_0 (p) dp\pm R
\end{equation}
where $Q_0 (p)$ is the initial distribution. Of course, this result is not exact as we have not weighted the averages using the cluster sizes.

\section{Conclusions}
\label{sec:conclusions}

Two simple extensions of the model of Egu\'{\i}luz and Zimmermann \cite{eguiluz} have been proposed to model the generation of the demand mechanism. In the original model, the level of transaction is arbitrarily fixed to a given value of $a$, with a transaction happening every $1/a$ time steps. In our models, agents are given a parameter $p$ that crudely models their characters, or more precisely, the way they are perceived in the market. Two agents of similar characters are supposed to get on quite well and to be able to make temporary associations. 

We distinguish between dictatorship-like associations, where a group of agents rely on one of them to make decision, and democratic-like associations, where the group decision reflects an average of the characters of all the agents belonging to the group. Both versions were investigated numerically and analytically, showing a very reasonable agreement between the two types of results. It is predicted and observed that agents in a democratic-like model make more durable associations. As these associations are groups of agents making common decisions, it results in very abrupt changes in the supply and demand balance. The resulting price, whose variations are defined through the balance of demand and supply by Eq. (\ref{eq:definition of the price}), has a very high volatility. This is what is known as the herding effect. On the contrary, in the dictatorship model, associations of agents are short-lived, giving less large groups of agents and a lower volatility in the price.

Both models exhibit organization. The democratic model is driven into a critical state where the average size of groups of agents is diverging. Associating a group of agents buying or selling with an avalanche, it is easy to show that the democratic model is self-organized, the size distribution of the avalanches being a power law. A separation between the time scales of the connections between agents and the separation of groups of agents is observed. The dictatorship model also exhibits a level of organization, with the spontaneous segregation of the agents into two distinct populations. Agents in the same population get on very well, making associations together. On the contrary, agents in different populations are very unlikely to connect. When two such groups of agents meet, it very often results in the separation into independent agents of one of the two groups. Because of the high probability of this fragmentation process, large associations of agents are very unlikely to form. The dictatorship model is not self-organized because it can never reach a critical state.

These models are more elaborate versions of the original models, because we attempt to model the demand process. In the original models, the fragmentation rate had to be tuned to a low value to ensure the appearance of groups of agents of all sizes and reproduce the herd effect observed in financial markets. We showed here that the fragmentation rate could be low because of democratic types of associations. Perhaps giving everybody their say on the management of portfolios is not the best way to ensure less volatile markets. Nevertheless, there remain still numerous rooms for improvements in the models. For instance, the buying or selling decision is made at random which is very unlikely to be realistic. Further studies along these lines could concentrate on including agents that analyse the returns before making a decision, as in \cite{dhulst2}.

Variations of the democratic rule have also been investigated, for instance weighting the value of $p$ after connection of two clusters according to the cluster sizes. The definition of the range $R$ has also been modified, choosing a range that is proportional to the size of the cluster that makes a transaction. Instead of breaking at random one of the two clusters when they are not connected, we also investigated the possibility that it is always the cluster with the greater $p$ that is fragmented. All these variations produced results that were qualitatively the same as one of our two models. Consequently, our results are robust and representative of this class of models.

\end{document}